\documentclass[prl,aps,twocolumn,epsfig]{revtex4}
\usepackage{epsfig}
\usepackage{graphicx}
\usepackage{amsmath}
\usepackage{amssymb}
\usepackage{bm}
\usepackage{times}

\newcommand{\be}{\begin{equation}}
\newcommand{\ee}{\end{equation}}
\newcommand{\beq}{\begin{equation}}
\newcommand{\eeq}{\end{equation}}
\newcommand{\bea}{\begin{eqnarray}}
\newcommand{\eea}{\end{eqnarray}}
\newcommand{\bei}{\begin{itemize}}
\newcommand{\eei}{\end{itemize}}

\def \r {{\bf r}}

\begin{document}

\title{Composite Fermion Theory for Bosonic Atoms in Optical Lattices}

\author{G.~M\"{o}ller$^1$ and N.~R.~Cooper$^{1,2}$}
\affiliation{$^1$Theory of Condensed Matter Group, Cavendish Laboratory, J.~J.~Thomson Ave., Cambridge CB3~0HE, UK\\
$^2$Laboratoire de Physique Th\'eorique et Mod\`eles Statistiques, 91406 Orsay, France}

\begin{abstract}
  We study the groundstates of cold atomic gases on rotating
  optical lattices, as described by the Bose-Hubbard model in a
  uniform effective magnetic field. Mapping the bosons to
  composite fermions leads to the prediction of quantum Hall fluids
  that have no counterpart in the continuum. We construct trial
  wavefunctions for these phases, and perform numerical tests of the
  predictions of the composite fermion model. Our results establish
  the existence of strongly correlated phases beyond those in the
  continuum limit, and provide evidence for a wider scope of the 
  composite fermion approach beyond its application to the lowest
  Landau-level.
\end{abstract}
\date{April 20, 2009}
\maketitle

Ultra-cold atomic gases have become a very active field of study of strongly
correlated quantum systems. While dilute Bose gases are typically in a weakly
interacting regime, they can be driven into regimes of strong correlations by
various means.  The application of an optical lattice potential leads to a
suppression of the kinetic energy relative to the interaction energy, and has
allowed the experimental exploration of the quantum phase transition between
Mott insulator and superfluid \cite{blochNature}.  Rapid rotation of the atomic gas
also leads to a quenching of the kinetic energy, into degenerate Landau
levels \cite{wgs}, and a regime of strong interactions \cite{schweikhard}. At
low filling factor $\nu$ (defined as the ratio of the number of particles to the 
number of vortices) this is predicted to lead to very interesting
strongly correlated phases \cite{cwg} which can be viewed as bosonic versions
of fractional quantum Hall effect (FQHE) states \cite{CooperReview}.
In order to access the low filling factor regime in experiment, it may be
favourable to exploit the strong interactions that are available in optical
lattice systems \cite{tung,hafezi} for which methods exist in which to simulate
uniform rotation (or equivalently a uniform magnetic
field) \cite{jaksch,mueller,sorensen}.  This raises the interesting question:
what are the correlated phases of atomic gases that are subjected both to an
optical lattice and to rapid rotation?

In this Letter, we study the interplay between the FQHE of bosons and the
strong correlation imposed by an optical lattice potential.  At sufficiently
low particle density, the effect of the lattice has been shown to have
negligible impact on the nature of the continuum Laughlin state at
$\nu=\frac{1}{2}$ \cite{sorensen,hafezi}. We focus on the possibility that
there exist strongly correlated phases which have no counterpart in the
continuum, but that appear as a direct consequence of both the lattice
potential and rotation.  To do so, we adapt the composite fermion (CF)
theory \cite{jain89,heinonen} which has been shown to accurately describe
rotating atomic Bose gases in the
continuum \cite{CooperWilkin99,regnaultJolicoeur03}, and apply this theory to
rotating bosons on a lattice.  Within mean-field theory, the lattice leads to
the intricate Hofstadter spectrum for the composite fermions \cite{kolRead}. We
predict a series of incompressible phases of bosons on the optical lattice,
characterized by special relations of the flux density $n_\phi$ and particle
density $n$, and we construct trial wavefunctions describing these phases.  From extensive exact diagonalization studies, we
establish the accuracy of the composite fermion approach, notably for states
for which $n=\frac{1}{2}\pm\frac{1}{2}n_\phi$; these correspond to
incompressible quantum Hall states which have no counterpart in the continuum.
To our knowledge, there has been no previous evidence for new FQHE states induced
by a lattice potential.
A previous proposal for quantum Hall states of bosons on the
lattice \cite{PalmerJaksch} takes a different viewpoint, but remains untested.

We study a model of bosonic atoms on a two dimensional square lattice and
subjected to a uniform effective magnetic field, using the Bose-Hubbard
model with Hamiltonian \cite{jaksch,mueller,sorensen}
\be
\label{eq:BoseHubbardH}
H = -J \sum_{\langle i,j\rangle } \left[\hat{a}_i^\dag\hat{a}_j e^{i A_{ij}} + h.c.\right] 
  + \frac{U}{2}\sum_i \hat{n}_i(\hat{n}_i-1),
\ee
with $\hat{a}^{(\dag)}_i$ bosonic field operators on site $i$, 
and $\hat{n}_i\equiv \hat{a}^\dag_i\hat{a}_i$.
We consider a uniform system with fixed average particle density $n$ (per
lattice site).  The strength of the magnetic field is set by the flux density
$n_\phi$ (per plaquette), defined by the condition that $\sum_\square A_{ij} =
2\pi n_\phi$.  Owing to the periodicity under $n_\phi\to n_\phi+1$, we choose
$0\leq n_\phi < 1$.

The single particle spectrum follows from the solution
of Harper's equation, and takes an intricate form, often referred to as the
Hofstadter butterfly \cite{hofstadter76}. It has a fractal structure consisting
of $q$ bands at rational flux density $n_\phi=p/q$. Signatures of this
structure appear in the mean-field treatment of the Bose-Hubbard
model \cite{oktel:045133,goldbaum:033629}. We wish to determine the groundstates of bosons
beyond the mean-field regime, where interparticle repulsion leads to
strongly correlated phases.
We focus on the hard-core limit $U\gg J$, where the bosonic Hilbert-space is
reduced to single occupations of lattice sites $0\leq n_i\leq 1$.
In this limit, using standard substitutions, the Hamiltonian
(\ref{eq:BoseHubbardH}) can be viewed as a spin-1/2 quantum magnet.
The gauge fields introduce frustration, putting this in the
class of frustrated quantum spin models where unconventional
spin-liquid phases can appear.
Indeed the Laughlin $\nu=\frac{1}{2}$ state studied in Ref.~\cite{sorensen} is
in the Kalmeyer-Laughlin \cite{kalmeyerlaughlin} spin-liquid
phase \footnote{Time-reversal symmetry is broken explicitly in the model.}. The
strongly correlated phases that we describe here can be viewed as
generalizations of this spin-liquid phase.

Following the application of composite fermion theory for rotating bosons in
the continuum \cite{CooperWilkin99}, we construct composite fermions by
attaching a single vortex to each boson. (An explicit form for the
wavefunction is presented below.)
The CF transformation relates the flux density for the original
atoms $n_\phi$ and the effective flux for CFs $n_\phi^*$ via
\be
\label{eq:CFtransform}
  n_\phi^* = n_\phi \pm n,
\ee
where the two signs correspond to attaching vortices of opposite sign.  Within
a mean-field theory, the CFs are assumed to be weakly interacting, and to form
a Fermi-sea which fills the lowest energy states of the single-particle
spectrum. Incompressible states then occur when the CFs completely fill an
integer number of bands. In the continuum, the single-particle spectrum
consists of Landau-levels (LL), leading to an incompressible state each time
an integer number, $\nu^*=n/n_\phi^*$, of CF Landau-levels is filled
\cite{CooperWilkin99,regnaultJolicoeur03}.  Applying the same logic on the lattice, leads to the conclusion that the single particle spectrum of the CFs is the Hofstadter
butterfly \cite{kolRead}, now at a flux density $n^*_\phi$. Owing to the
fractal structure of this energy spectrum, depending on $n^*_\phi$ there can
be many such energy gaps, leading to many possible incompressible states.
To determine the locations of these incompressible states, we need to know the
particle densities $n$ which completely fill an energy number of bands of the
spectrum of CF's at flux $n_\phi^*$.  Generalizing from the continuum where
the density of states in each LL is proportional to the flux density, an
analysis of the spectrum leads to the conclusion that, when filling all states
up to any given gap in the Hofstadter spectrum, the relation between $n$ and
$n_\phi^*$ remains linear \cite{claroWannier,usCondensed}, 
$
  n = \nu^* n_\phi^* + \delta,
$
with an offset $\delta$. The coefficients $\nu^*$ and $\delta$
can be determined from the Hofstadter spectrum by locating two points
within the same gap.
\begin{figure}[ttp]
\includegraphics[width=1.0\columnwidth]{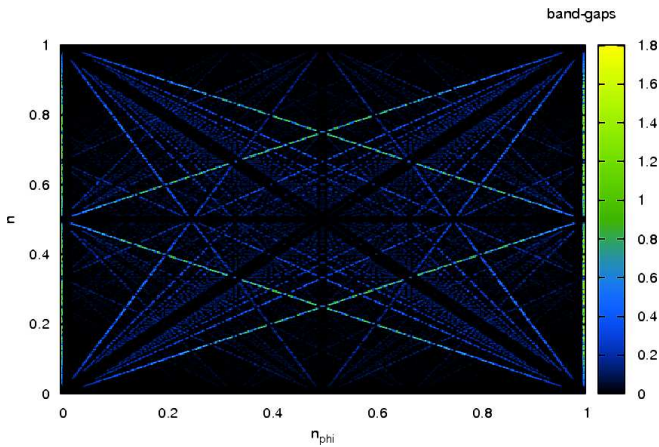}
\caption{ \label{fig:gapsCFs} Excitation gap of bosons
on a lattice, with particle density $n$ and flux density $n_\phi$, as
predicted by a model of non-interacting composite fermions.  The
bright lines show parameters $(n,n_\phi)$ where the model predicts the
appearance of incompressible quantum fluids, and include cases (where
$n/n_\phi$ is not constant) which are not connected to the continuum
limit.  We include data for 
$n_\phi^*=p/q$, with $q\leq 50$.}
\end{figure}
Then, using the reverse of the CF transformation (\ref{eq:CFtransform}), one
obtains the lines of $n,\,n_\phi$ on which a non-zero gap is predicted above
the CF groundstate.

Within a model of non-interacting CFs the relative magnitudes of gaps follow
from those in the single particle CF spectrum. The gaps inferred under this
hypothesis are shown in Fig.~\ref{fig:gapsCFs}, in which the mode of flux
attachment [corresponding to the positive and negative sign in
(\ref{eq:CFtransform})] is chosen to give the maximum gap.  Note that the
positive sign in (\ref{eq:CFtransform}) can be regarded either as negative
flux attachment \cite{MollerSimon05}, or as attachment of the conjugate flux
$1-n$ due to the particle-hole symmetry on the lattice.  Indeed,
Fig.~\ref{fig:gapsCFs} shows symmetries under $n_\phi \leftrightarrow 1-n_\phi$ and $n \leftrightarrow
1-n$. In the hardcore limit, the Hamiltonian itself enjoys these
symmetries, so the parameter space may be reduced to $0\leq n,n_\phi\leq
\frac{1}{2}$.
In this quadrant, the lines emerging from
the corner with $n=n_\phi= 0$ and constant filling factor $\nu \equiv n/n_\phi$
are the CF states expected in the continuum
limit \cite{CooperWilkin99,regnaultJolicoeur03}.  Crucially, however,
Fig.~\ref{fig:gapsCFs} shows a large number of other lines.  These correspond
to new candidate incompressible states. 

The preceding discussion conjectures candidates for new types of
correlated quantum liquids of bosons in optical lattices.  However, given that
the mean-field CF theory is an uncontrolled approximation, it is
important to test these predictions.  There are competing condensed
states on the lattice \cite{usCondensed,PalmerJakschLong,kasamatsu}.  Even in the
continuum limit, some of the correlated states predicted by composite
fermion theory are replaced by other strongly correlated phases
\cite{wgs}, with only $\nu=1/2, 2/3$ and $3/4$ appearing to be described in
this form \cite{regnaultJolicoeur03}.

We have investigated the success of the CF construction for the Bose-Hubbard
model (\ref{eq:BoseHubbardH}) using exact diagonalisation studies.
We study the model for $N$ particles on a square lattice with
$N_s=L_x\times L_y$ sites, in the presence of $0\leq N_\phi < N_s$ flux quanta. To limit finite size effects, we impose periodic boundary conditions (pbc)
giving the system the geometry of a torus. (The nature of these boundary conditions is discussed further
below.)  

In order to compare the exact groundstates with the CF theory it is
useful to have a trial CF wavefunction. We generalize the continuum
construction \cite{CooperWilkin99} to allow not only for the lattice,
but also for the torus geometry, for which no convenient formation
exists even in the continuum limit.  We construct the trial CF state
for bosons in a lattice, by writing
\be
\label{eq:CFState}
\Psi_\text{trial}(\r_1,\ldots,\r_N) = \Psi_\text{J} (\{\r_i\})
\times \Psi_\text{CF}  (\{\r_i\})
\ee  
where $\Psi_\text{J}$ and $\Psi_\text{CF}$ are {\it fermionic} wavefunctions.
The factor $\Psi_\text{J}$ effects the flux attachment (\ref{eq:CFtransform}),
and represents the ``Jastrow'' factor of the continuum
wavefunction \cite{CooperWilkin99}, but is now defined on a lattice and in the
torus geometry. 
Instead of using the continuum form \cite{readRezayi96}, we note that this 
function corresponds to a filled Landau-level of fermions at flux $\mp N$, 
and we generate $\Psi_\text{J}$ on the
lattice as the Slater determinant describing $N$ fermions occupying the $N$
lowest energy states on the lattice with $\mp N$ flux. The factor
$\Psi_\text{CF}$ is the wavefunction of the CFs in the resulting effective
field (\ref{eq:CFtransform}). This is described by the Slater determinant of
the $N$ lowest energy single-particle CF states $\Phi_k$ at flux
$N_\phi^*=N_\phi\pm N$. For the cases derived above (and illustrated in
Fig.~\ref{fig:gapsCFs}), these numbers $N$ and $N^*_\phi$ are such that the CFs
fill a integer number of bands.  Note that, in contrast to the continuum
limit where the groundstates have been studied within the lowest LL
limit \cite{CooperReview}, our CF state (\ref{eq:CFState}) does \emph{not}
include a projection to the lowest LL.  This is appropriate for the hard-core
model that we study, since (\ref{eq:CFState}) vanishes when
the positions of any two bosons coincide.

The description of the trial state (\ref{eq:CFState}) is completed by
discussing the boundary conditions imposed on each of the functions.
In the most general case, one introduces twisted boundary conditions
for the bosons, defined by the phases $\theta_\mu=(\theta_x,\theta_y)$, such 
that magnetic translations of a boson around the two cycles of the torus (by 
$L_\mu$ in the $\mu$-direction) act as $\Psi \to \exp[i \theta_{\mu}]\Psi$.
In the definition of the trial state (\ref{eq:CFState}), one may
choose boundary phases for the Jastrow- and CF-parts independently,
defining $\theta_\mu^\text{J}$, $\theta_\mu^\text{CF}$.  The sum of
these phases is constrained to match the pbc for the bosons
$\theta_\mu^\text{J}+\theta_\mu^\text{CF}=\theta_\mu$.  However, this
still leaves the freedom to vary
$\theta_\mu^\text{J}-\theta_\mu^\text{CF}$. 
This freedom is a crucial
ingredient to our construction: it allows one to generate the set of
states responsible for the non-trivial groundstate degeneracy of these
topologically ordered phases on a torus \cite{Wen}.  It is
easy to show that, with this freedom, in the continuum limit the
wavefunctions (\ref{eq:CFState}) reproduce the two continuum
Laughlin states at $\nu=1/2$ \cite{HaldaneRezayiTorus}.

As an initial test, we have computed the overlaps $|\langle
\Psi_\text{trial}|\Psi_\text{exact}\rangle|^2$ of our trial wavefunctions
(\ref{eq:CFState}) with the groundstates on the lattice at $\nu=1/2$ as
a function of $n_\phi$.  The overlaps (not shown) are very close to those found 
with the continuum Laughlin wavefunctions \cite{sorensen}.  Thus,
there is little difference between the continuum \cite{HaldaneRezayiTorus} and
lattice state (\ref{eq:CFState}), up to the flux density
$n_\phi\simeq 0.3$ at which the overlap falls to a small value.

\begin{figure}[ttp]
\includegraphics[width=0.99\columnwidth]{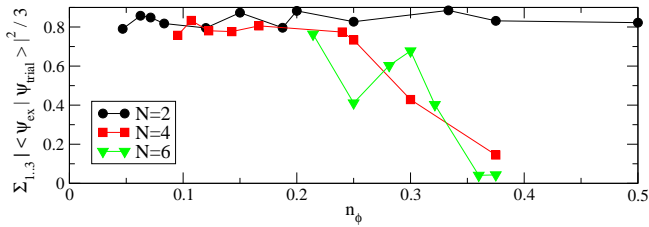}
\caption{ \label{fig:overlapsFQHE} Average overlap of the CF states with the 
exact eigenstates in the (approximately) three-fold degenerate groundstate manifold
of the $\nu=\frac{2}{3}$ state. }
\end{figure}

Using our general construction (\ref{eq:CFState}), we are able to
study for the first time the influence of the lattice structure on
other continuum CF states. The state at $\nu=2/3$ has a groundstate
degeneracy $d_{\rm GS}=3$ \cite{HaldaneRezayiTorus}. We take overlaps
of the CF trial states within the groundstate manifold composed of the
three lowest states of the exact spectrum, and give their average
value in Fig.~\ref{fig:overlapsFQHE}.  The overlap is high and drops
only above flux densities of $n_\phi\simeq 0.35$.  Previous numerical
evidence for the appearance of this CF state is restricted to the
lowest Landau level \cite{CooperReview}. Our results show that, for
sufficiently small $n_\phi$, the CF state (\ref{eq:CFState}) also
describes the groundstate for hard-core interactions (where LL mixing
is strong).

Let us now return to the main focus of this Letter: the new
CF states that appear on the lattice. To investigate these
states numerically
we focus on the CF series derived from the most dominant gap in a
subcell of the Hofstadter spectrum (cell $L_1$ \cite{hofstadter76}),
leading to a sequence with $n_\phi = \frac{1}{2}-\frac{1}{2}n$.
In order to be able to study several different system sizes for some states in
this class, we select two points where ($n,n_\phi$) are fractions with
small denominators, and the particle density is low enough to avoid competition
with the continuum Laughlin state: $(\frac{1}{7},\frac{3}{7})$, and 
$(n,n_\phi)=(\frac{1}{9},\frac{4}{9})$. 
\begin{table}
  \begin{center}
    \begin{tabular}{cccccccccccccccccc}
      $n$ &&$n_\phi$ &&$N$ &&$L_x$ &&$L_y$ &&$\Delta$ &&$\mathcal{O}_{\rm CF}$ &&$\dim(\mathcal{H})$\\
      \hline
      1/7 &&3/7 &&2 &&2 &&7 &&0.156 &&0.437 &&91\\
      1/7 &&3/7 &&3 &&3 &&7 &&0.156 &&0.745 &&1330\\
      1/7 &&3/7 &&4 &&4 &&7 &&-0.032 &&0.2753$^\dagger$ &&20.5k\\
      1/7 &&3/7 &&5 &&5 &&7 &&0.0401 &&0.5631 &&324k\\
      1/7 &&3/7 &&6 &&6 &&7 &&0.0455 &&0.3284 &&5.2M\\
      \hline
      1/9 &&4/9 &&2 &&2 &&9 &&0.113 &&0.3603 &&153\\
      1/9 &&4/9 &&3 &&3 &&9 &&0.241 &&0.8407 &&2925\\
      1/9 &&4/9 &&4 &&4 &&9 &&-0.036 &&0.1515$^\dagger$ &&58.9k\\
      1/9 &&4/9 &&4 &&6 &&6 &&0.071 &&0.3061 &&58.9k\\
      1/9 &&4/9 &&5 &&5 &&9 &&0.0945 &&0.4585 &&1.2M\\
      1/9 &&4/9 &&6 &&6 &&9 &&-0.0154 &&0.1957$^\dagger$ &&25.8M\\
      \hline\hline
    \end{tabular}
  \end{center}
  
  \caption{\label{tab:CFdata}
    Exact diagonalization results: gaps $\Delta$ and overlaps 
    of the exact groundstate with the CF trial state with negative flux attachment
    $\mathcal{O}_\text{CF}=|\langle \Psi_\text{trial}|\Psi_\text{ex}\rangle|^2$.
    We also give the Hilbert-space dimension $\dim(\mathcal{H})$ for hardcore bosons. 
    Where negative gaps are indicated, the CF state is the first excited state.  
    Notes: $^\dag$overlap shown for first excited state.
    }
\end{table}

We find multiple pieces of evidence for the formation of strongly correlated
incompressible phases at these values of $(n,n_\phi)$.  
First, an analysis of the eigenvalues of the single particle density
matrix of the groundstate shows that, as the system size $N$
increases, there are $N$ eigenvalues of order one. Thus, there is no
evidence for condensation (an eigenvalue that grows with $N$), so the
groundstate is likely uncondensed, and strongly correlated.
Second, the spectra at these densities typically show a single
groundstate separated by a large gap (see Table \ref{tab:CFdata}).
 The gaps we find are
larger than the typical spacing of higher excited states, or the gaps
at typical spectra at nearby flux densities \footnote{Given the limited
number of data-points, we do not attempt a finite-size scaling of the
gaps.}.
This indicates that the system may be an incompressible liquid with a
non-degenerate groundstate on the torus.
This is consistent with the
CF state, in which one expects a groundstate degeneracy of one,
applying the reasoning of Ref.~\cite{kolRead}.

Further direct evidence for the CF phase is obtained by taking the
overlap of the exact groundstates with the trial CF states
(\ref{eq:CFState}).
As detailed in Table \ref{tab:CFdata}, we find that, in general, the
trial CF states have significant overlap with the exact groundstate.
Notable exceptions occur for certain cases ($N=4$ for $n=1/7$ and $n=1/9$, 
and $N=6$ for $n=1/9$) where
the exact groundstate has a different momentum from the CF state so
the overlap vanishes identically. In these cases, we find large
overlap of the CF state with the lowest lying excited state (as shown
in Table~\ref{tab:CFdata}). We account for this behaviour as arising from the
existence of a competing broken-symmetry ``stripe'' phase that is
stabilized by delocalization of the particles around the short
direction, similar to finite size effects in continuum
studies on the torus \cite{CooperRezayi}. This interpretation is
confirmed by our studies at $n=1/9$, which show that the groundstate
is sensitive to the lattice geometry (two aspect ratios $L_x\times
L_y=4\times 9$ and $6\times 6$ are available for $N=4$ at $n=1/9$).
The groundstate reverts to be of the CF form for the more
isotropic aspect ratio.
Unfortunately, no geometry with smaller aspect ratio is available for 
the systems ($N=4$ at $n=1/7$ and $N=6$ at $n=1/9$).
Still, our results indicate that, for the system at 
$(n,n_\phi)=(\frac{1}{7},\frac{3}{7})$, the composite fermion state 
dominates the competing (striped) state at large system sizes, and 
maintains a very high overlap with the exact groundstate.
A similar trend is evident for $(n,n_\phi)=(\frac{1}{9},\frac{4}{9})$, 
but in this case the available geometries at $N=6$ are still very asymmetric 
so we cannot confirm the preference of the CF state in this case.

Overall, the values of the overlaps with the CF state are highly
non-trivial, given the sizes of the Hilbert spaces and that the trial
CF wavefunction has no free parameters. In contrast, for large system
sizes, the overlap with a condensed (Gutzwiller) wavefunction is much
smaller ($< 10\%$), even allowing for optimization over the condensate
wavefunction. It is clear that the CF ansatz is capturing the
essential physics of the correlated phases.

While a large overlap with the trial CF state is highly suggestive
that the phase is of the CF type, it is very useful to have other
tests of the {\it qualitative} features of the state.  As noted above,
the nondegenerate groundstate is consistent with the expected
topological degeneracy of the CF state.  Another important qualitative
test is provided by Chern numbers \cite{Hatsugai,hafezi}, which
provide a highly non-trivial test of the existence and nature of the
topological order of a many-body quantum phase.  We have evaluated the
Chern number $\mathcal{C}$ for the groundstates with nonzero overlap
with the CF states for systems up to $N=5$. In all cases, we find that
$\mathcal{C}=2$. This is the value expected for the CF phase
\cite{kolRead}. This agreement lends very strong evidence that the
phase appearing in the numerics is of the form predicted by the CF
theory.

In conclusion, we have presented numerical evidence for novel types of
correlated quantum fluids for bosons on rotating lattices. 
Our results show strong evidence that there 
exist fractional quantum Hall states beyond those in the continuum limit. They provide the first
evidence for a wider applicability of the composite fermion ansatz. They also further
motivate experimental studies of rotating gas on optical lattices, which
would be able to probe novel aspects of the physics of quantum Hall systems.

\acknowledgments{
GM acknowledges support from ICAM and
the hospitality of the University of Colorado.
This work was supported by EPSRC Grant
No. GR/S61263/01 and by IFRAF. }

\bibliography{latticeCF}

\end{document}